# Danse-doigts, a Fine Motor Game

*Jean-Ferdy Susini1, *Olivier Pons, **Nolwenn Guedin,***Catherine Thevenot

*CNAM, CÉDRIC,292 rue Saint-Martin 75141 paris cédex 03
(jean-ferdinand.susini@cnam.fr; olivier.pons@cnam.fr)

**FPSE, UNIGE,  40 bd du Pont d'Arve, 1205, Genève 4, Switzerland

(nolwenn.guedin@unige.ch)

***IP, UNIL, Géopolis - 1015 Lausanne,  Switzerland

(catherine.thevenot@unil.ch)

## Abstract

This paper describes the design, implementation and testing of "*Danse-doigts*", an edutainment therapeutic application for hemiplegic children. The objective of this program is twofold. Firstly, to allow them to train their fine motor skills on tablet. Secondly, to study the effect of this training on their numerical performance (counting, calculation...). The target population and the objective of evaluating numerical skills influenced the design. The software was developed using standard web technologies but is based on a new parallel programming library written in JavaScript.

Applications and libraries are free of charge and easy to install on most tablets.

## 1. Introduction

In this article we present the approach used during the implementation of "*Danse-doigts*", an edutainment therapeutic application for the rehabilitation of the upper limb of hemiparetic children. Our goal is twofold. Firstly to provide a training tool for hand fine motor skills. Secondly, to measure the impact of this training on numerical skills (counting, calculation...).

This duality makes the design of "*Danse-doigt*" particularly interesting to study. Indeed, we tried to design a game whose playful and ergonomic aspects are relevant to the target population, but also to incorporate finely therapeutic intentions, without disrupting the intended study.

The lack of accurate data on the target population (degree of motor impairment, presence and extent of spasticity, presence of associated disorders but also morphological characteristics such as size of the hand) encouraged a modular and adaptive design. The use of specific programming libraries, which we present in Section 5, greatly facilitated this design.

We made assumptions based on our own intuition and experience. We then revised them in accordance with the incremental feedback of the experiments carried out in vivo by therapists.

The next section is a short review of some applications and games involving motor skills of the hand. In the third section we present the context and the specifications. Section four specifies the assumptions and methodology used to create playful mechanism, accessible and motivating for an audience of children, while remaining compatible with the constraints of the specifications. Section five describes the main features of the framework and programming libraries developed [1]. Section six presents the results of a preliminary experiment conducted on two classes of schoolchildren (without disabilities). On one side, this study allowed to adapt the software to the real constraints. On the other side, it aimed at investigating the relationship between numbers and fingers. We conclude this study with a discussion of the results of the experiment, a generalization of our approach and some proposed extensions.

## 2. Existing tools

Applications to train fine motor skills on tablet are numerous, but many of them also involve a cognitive work (puzzle, pre-writing skills...). Those targeting only motor function are rarely configurable and often inaccessible for a player with upper limb motor impairment.

Among the most used applications in motor rehabilitation, two seem to stand out:

1. Fruit Ninja[2], developed by Halfbrick where you have to cut fruits by swiping your touch screen with fingers while avoiding cutting bombs. This is now a classic, which has been ported to all platforms and is available in different versions. He was initially not therapeutic but was widely used for this purpose [3, 4].

2. The Dexteria suite, developed by BinaryLabs includes "Dexteria Fine Motor Skill Development" [5] which claims to be :« A set of therapeutic hand exercises (not games) to improve fine motor skills and pre-writing among children and adults. ».

Dexteria exercises take advantage of the multi-touch interface to train various gesture and aim to improve strength, control, and dexterity.

However if their different Game Play provide inspiration, none of these applications seem to have the characteristics what are relevant to us, in terms of software development, to allow optimal use of rehabilitation applications:
– To be easily deployable at lower costs (without rewriting) on any kind of tablet
– To be easily customizable and configurable according to the therapeutic or experimental constraints.
– To allow to closely monitor all interactions with the "user" at different levels of accuracy.
– To be extensible, allowing to add new features without compromising the existing.

The development platform must also be open source. Thus any research team would be allowed to tailor it in order to meet their particular needs.

## 3. Constraints of the specifications

We now present the target population and the objective of the planned study. We also study their influence on the design of the application.

## 3.1 The Target: the hemiparetic child

Hemiparesis is an unilateral paresis due to brain injury. Relatively common in children, with a prevalence of 1/1,000, it may be congenital or acquired. It results in various level and kind of incapacity and may be associated with other disorders (epilepsy, spatial and visual disturbances...). This diversity complicates the specification and design of specific rehabilitation applications. We only target the upper limb. In the first version of the application (dedicated to the study), we only assumed a sufficient mobility to click with different fingers. In the next versions we also assumed an (even limited) ability to pinch (thumb-index oppositions) which is one of the main difficulty for a hemiplegic hand.

Moreover, in addition to school activities a child with motor impairments is faced with a lot of regular therapeutic appointments (physiotherapy, occupational therapy, psychomotricity...). A training that complements traditional therapies must not too overloaded. Thus, it should be done at home or at school. This reinforces the imperatives of cost, deployability and ease of use already mentioned in Section 2. This also requires introducing clear guidelines in the GamePlay.

## 3.2 Objective of the study: fingers and numerical abilities

From the early years of learning and sometimes even into adulthood, we use our fingers to represent and communicate quantities and to perform calculations [6]. Described in 1940, the Gerstmann syndrom [7] suggests a link between impaired finger gnosia and low numerical skills. More recent experimental studies investigated the nature of the relationship between finger abilities and numerical performance. They have shown that the relationship between finger use and numerical representations might be related to the anatomical proximity of relevant brain areas [8-10]. In addition to this neuronal explanation, other works tend to explain this link by the early finger use in functional motor habits like pointing objects while counting, or calculating on fingers. These motor recruitments may help the good development of numerical skills [11,12]. The finger gnosia task assesses the tactile ability to recognize our own fingers without visual cues. Compared to classical neuropsychological tests, this sensorimotor task performed by five-year old children is a better predictor for their mathematical achievement one year latter [13]. Moreover, the spontaneous fingers use in counting tasks increases from the age of 4 until 7 and is correlated with a better success rate [14]. In most cases, low finger gnosia is associated to mathematical troubles even in children from 8 to 11 years [12, 15, 16]. However, some children with bad fingers gnosia because of their congenital hemiparesis are able to sufficiently compensate their initial numerical difficulties to finally reach the same arithmetical level as children with typical development. [17].

If finger abilities are linked with numerical activity, training them should lead to improvements also in mathematical tasks. Proposed actions have to stimulate fingers one by one in order to study the impact of such training in numerical tasks. To do this, we have chosen pointing actions that should be done successively with each finger from both hands. To date, only one study assessed this hypothesis with children 6 to 8 years [18]. In this study, it was asked to point target and to follow mazes 30 minutes a day for 8 weeks. It showed improvement in mathematical performance, but the validity of these results has been questioned [19].

One of the main objective of the study conducted with "*Danse-doigts*" is to try to confirm these results and later to study the gains for children with hemiparesis.

### 3.3 Design constraints

We now present the constraints (sometimes contradictory) that guided the design.

### 3.3.1 Maintain the motivation

Rehabilitation activity has to be steady and continued over the long term. This can quickly become repetitive, unrewarding and tedious, resulting in demotivation of patients who are then likely to follow the protocol irregularly or even to abandon it. Game design methodologies and gamification generally propose several elements likely to maintain engagement. The main ones are: storylines, goals directed around targeted skills, rewards and feedback about goal progress, increasing levels of difficulty, individualized training, and the provision of choice [20, 21].

### 3.3.2 Avoid circumvention

Motivated by the game, the patient may be tempted to divert the protocol. It can, for example, always use the index in pointing tasks or, in the case of a hemiparetic patient, using his good hand when asked to use their paretic hand. It is essential to try at least to detect these diversions and to prevent them by adding the necessary constraints to the *GamePlay*. To ensure the use of a specific hand on a task, we can, for example, use the other hand in parallel for a static task as in Figure 1.

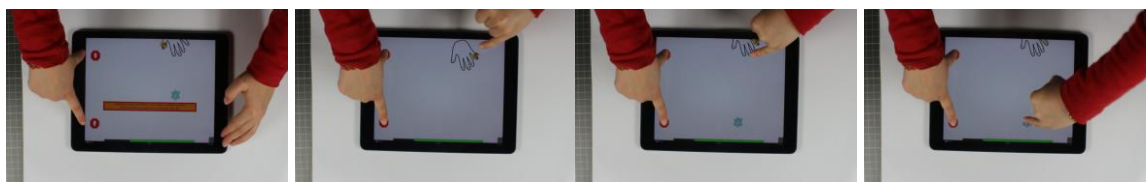

FIG. 1: Two fingers of the hand that does not play should remain placed on the "stop" signs. A small hand with a crown, indicating which finger should play, will be displayed. The child clicks on the crown with the required finger (here the little finger) to display the target. When the child clicks on the target (or if the timeout is exceeded), we restart with another finger.

### 3.3.3 Avoid any bias

Since we want to study the influence of finger training on numerical skills, in order to avoid any bias, software should not use any numerical information. We cannot count points, successes or failures. We neither have to numerically estimate the number of problems or the time remaining. It was also necessary to avoid any cognitive bias such a diversion of attention (background music, parasitic event, graphic changing environment ...)

## 4. « Game Design »

## 4.1 Reconciling constraints

The software designed for the numerical study is a fun application requiring only to point different targets using different fingers. The constraints expressed in Section 3.3 greatly reduce the fun side. We have nevertheless tried to maintain it by writing a story around repetitive tasks. Several themes are offered each containing several sub-themes. Child have to play some time before to discover a complete theme (one week in the experiment described in section 6). For example, in the nature topic for one week, children had five sorts of sub-topic, each for one day of training: "the sky", "the farm", "the garden", "the forest" and "the lake". In one-day ten-minute session, the child has to do four different games, each during two minutes and an half. For example, the training "the sky" includes the games "clouds", "flakes", "sun", rain" (different target sizes).

For each challenge, when child succeed, they hear a fun sound related to the target picture (audio reinforcement). Each session ends with a congratulation message and a recompense picture with a visual summary of the four games. For example, at the end of the sub-topic "sky", the child wins a beautiful picture with rain, sun, cloud and a snowman. During the first week of the experimentation (described Section 6) it appears that the removal of any reference to the remaining playing time embarrassed children so the introduced a progress bar to avoid numerical information. The constraints of the study also limit the motor ability that we try to train. The version dedicated to the experimentation, only use pointing tasks. Activities requiring several fingers, as pinching or grasping, were not used in this first experiment (exist in new version).

## 4.2 The game

When the child launches the application with the theme of the week, first appear the instructions to follow during the session. Then, he can choose a subtopic (eg. "Sky"). The 4 games that compose the subtheme then appear (eg., "Rain", "Sun" ...). He then chooses the one to start first. Two "stop" signs then appear on the screen, to the right or left according to the hand that should be trained. To continue the game, child must put, two fingers on the signs (Figure 1). They must stay on the signs during the game and the game stops if they are removed.

Once the fingers in contact with "stop" signs, a picture of the other hand appears at the bottom of the screen, one finger wearing a crown. Child must then click on the crown with his own (corresponding) finger. Next he use this finger to click target that is proposed. Targets may be static or moving and have different sizes. The selected finger changes between each target so the child must remain alert and focused to distinguish its various fingers at all times. Fingers selected to click (shown by the crown) were randomly chosen from a cycle of five fingers. There is no specific order, which limits potential anticipation. Thus, discrimination of the different fingers is done without habituation to a series of repeated actions. When clicked, the target disappears and another takes its place at a different position. This is always accompanied by a "fun" sound ( related to target picture). This combination of image and sound provides better stimulation of children because the effects of the observation are increased when the inputs are multimodal [22]. This allows better learning process.

## 5 Software Design and development

We now present some technical elements concerning the implementation of the application. We discuss in particular the programming libraries developed to ease the implementation of reactive game applications.

## 5.1 Technical implications

If the constraints 1 and 3 described in section 3.3 are mainly within the application design, we estimated that compliance with constraint 2 must be partly guaranteed by the application itself. The specification assumes that we have to continuously monitor the behavior of children's hands. Modulo some simple assumptions, detecting multiple points of contact on touchscreen allows, to some extent, to comply with this requirement.

However, during the experiment, our design assumptions are validated by visual inspection of an adult. Thus, to control the immobilized hand, our application monitors two points of contact "sufficiently" distant from each other. Those points are not in the screen area used by the game.

Working with a single targets, at evrery moment we should observe only one single point of contact corresponding to the finger clicking the target. Our application can detect multiple points of contact in the playing area and warn about non-compliance whith the instructions ("use only one finger to click"). However, in our experiments we only focused on the success of the exercise, ie detecting contact point within the target area. Other possible points of contact are ignored. But technically we could decide to record these unexpected contacts.

Notice that our application is also able to follow sliding touch points eg when performing exercises such pinching to merge two targets or to close fingers on a specific target.

To summarize, our application should be able to:

– continually monitor at least 3 contact points;
– detect any unexpected points of contact in the playing area;
– follow moving points of contacts in the playing area;
– manage graphical applications (displaying static targets, trajectories of moving targets, displaying instructions, progress bar ...)
– coordinate the broadcast of rewards sounds and audio instructions;
– observe all interactions to get statistical data.

This large number of activities to manage simultaneously (in parallel), led us to consider a parallel programming paradigm to ease implementation. As discussed in the subsection V-C, we chose the programming language JavaScript to write our application. JavaScript is present in any Web browser, but can not easily handle parallelism (in its standard version). So we used our SugarCubesJS framework that brings this dimension to JavaScript.

## 5.2 Parallelism and reactive programming library

The SugarCubesJS library allows to write programs designed as an aggregate of tasks executed in parallel. This library is based on the synchronous reactive programing model "à la" Boussinot. [23]

In short, this programming model proposes to split the execution of a computer system in parts of finite time executed one after the other. Time is considered logically as a sequence of (instantaneous) steps. This allows to see the execution as a succession of logical instants. This sequence is called a clock and drives the system execution. The concept of clock, make it possible to consider the program as a set of elementary tasks running in parallel in the same instant. In this programming model, synchronization of parallel tasks becomes a natural property. Indeed, the programming model itself supports the complexity of communication between tasks and the complexity of the synchronization. This makes programming very easy, simply write small independent tasks and then to put in parallel with each other.

According to this paradigm, the SugarCubesJS framework proposes a particular notion of events that differs from conventional Javascript events. An event is a global object (accessible by all parallel components) characterized by its presence or absence during an instant. At each instant of execution, an event is either present or absent. We say that these events are instantly broadcasted because they are accessible by all parallel tasks (broadcast) during the same instant (instantaneous). When present at an instant, an event can have a list of associated objects (list of value accessible by all interested components). The coherence of the state of a given event in a logical instant ensures that all components access the same list of values associated exactly at the same logical instant. The list of values is only accessible during the instant where it is defined. At each insant, a new list of values is be defined if the event is present. Interactions between parallel tasks are managed by the instantaneous broadcast of events.

During the development of "Danse-doigt", the use of this reactive paradigm has greatly facilitated the programming of parallel monitoring of contact points (some simple tasks in parallel can generate events reporting the supposed behavior of the player). It also make aesier to manage the combinatoric coming from the game logic, the animations and the sounds (audio reinforcement on success, instructions ...).

Another very important advantage, this facilitate the introduction of observers behaviors. It is thus very simple to add observer programs in parallel in a reactive system. They listen to the data sent by the initial system (observed system). Observers manage in parallel listened data to filter, analyze and store, without disrupting the system observed.

The collect of experimental data can be implanted with SugarCubesJS completely independently of the program managing the game and of tasks controlling the player behavior. The tasks collecting data are written as parallel activities simply listening to events emitted by the other tasks (manager of contact points, manager of game logic, animation,...). Depending on the expected statistical data, tasks that collect them may also perform aggregations, correlations and various other pretreatments on the data before transmission to the collect server. All this treatments never affects the development of the game or the interaction with the user. The SugarCubesJS tasks are independent of each other, our application can be run with or without the task collecting data. The tasks of the game broadcast their events without ever having "conscience" of the existence of other tasks running in parallel (technically, you never manipulate references to other tasks). The presence or absence of the collect task does not affects the behavior of the application. To illustrate this we pushed our experimentation to develop the collect tasks on an independent server.

## 5.3 « WebApp » development

To facilitate the deployment of the application on most tablets, we choosed a solution independent from systems, platforms and manufactureur. This solution is a "WebApp". It uses only HTML, CSS and Javascript, and provides easy access to the application (via a web browser). The well-established standards of the Web industry make easier to write applications that can be deployed without extra cost, on tablets used in families of children participating in the experiment.

The integration on Android and iOS system is enhanced through the use of a particular web browser one each platform: Chrome for Android and Safari for iOS. Indeed, after-loading the app's Web page in the browser, it can be installed on the tablet home screen, Creating a shortcut, simulating the behavior of a native application. When the user launches the app from this shortcut, the interface of our web application, is exposed in full screen, looking like a native application.

Besides the ease of access, this technical solution also makes it fairly easy to manage updates by correcting files directly on the web server. Our framework also allows to use the application outside internet connection. However, an Internet connection is still required for the collect of statistical data. In temporary absence of connection, our framework plans to save data in the tablet data, the time to regain network access.

## 6 Experience and extensions

A pilot study [24,25] of 9 weeks was led with children with typical development in first and second years of primary school. Two groups were constituted: a training group who played with tablets and a control group. Seventeen children were in each group and were equally distributed according to their finger gnosia level. The mean age was of 6.6 years. The training was done during school time, four times a week, with a session of 10 minutes per day. Children played in the computer room with audio headsets in order to follow oral instructions at individual rhythms.

Several tests were assessed before and after the training (Nine-Hole Peg Test, Finger gnosia task and evaluation of numerical skills). In the assessment of the finger gnosia, children have to identify which of their fingertips the experimenter touches. In our adaptation of the task, children wore fingerless gloves, with the three central fingers of each glove of different colors. Children were asked to place their hands palm down through a lateral aperture of a covered box. The examiner touched one fingertip through an opposite aperture and rapidly opened the cover of the box. Then, children had to name the color corresponding to the touched fingers.

Digital skills were evaluated on non-symbolic material (quick comparison of dot clouds) on semi-symbolic quantities (recognition of digital configurations) and finally on symbolic representations (additions). Statistical analyses with adequate ANOVA did not show any significant progress in numerical skills even not in finger abilities. This lack of improvement could be due to a shorter time of training.

We also experimented some extensions that have not yet been studied on children:
– Use two fingers (the thumb and an opposable finger) to pinch. The child must touch two targets (one finger on each target) and slide his fingers on the screen to merge target.
– A variation of previous exercise is to ask to the child to close the pinch on a static target.

- It is possible to complicate the previous exercise by transforming the static target into a moving target ( with a predictable trajectory )
- It is also very easy to increase the number of fingers simultaneously working. But the tablet must be able to detect all contact points. Most tablets recognize at least 5 points of contact (allowing to work with 3 fingers, two contacts being used by the immobilization constraint on the other hand). More modern tablets recognize a dozen contacts which is more than enough for this kind of exercise.
- Finally, we can easily perform pointing or sliding exercises whit 5 fingers (if the hardware actually supports)

Tracking all signals from the user action allows to analyze action with different levels of precision.

## 7 Discussion & Conclusion

Various versions of the application "*Danse-doigt*" [26] and the programming library SugarCubesJS[1] are available online.

The game was well received and seems easy to pick up and play for both children and supervisors. The experiment conducted in regular classes did not confirm the expected results. This may be due to too short a training time. However, the study should be continued with hemiplegic patients. Moreover, user feedback in the first week of use has helped to refine the constraints and to correct some design mistakes. These incremental updates helped to extensively test the framework. SugarCubesJS demonstrated that it is a mature framework easy to **maintain**, optimize, and extend.

We also want to study the effectiveness of training with our various variants of exercises using classical motor assessments scales. For this, the use of the SugarCubesJS library and the framework developed for the experiments allow to quickly develop new versions to train other skills or for a specific study.

## Thanks

We thank the organisation "Association Hémiparésie" who initiates the project and the staff and children of the Tivoli school in Dijon where experiments were conducted.